\documentstyle[12pt]{article}

\newcommand\beq{\begin{eqnarray}}
\newcommand\eeq{\end{eqnarray}}

\renewcommand{\theequation}{\thesection.\arabic{equation}}

\newcommand\la{\langle}
\newcommand\ra{\rangle}

\begin{document}
\hfill UTHEP-271

\hfill  January 1994

\begin{center}
\Large
\bf
QCD and Quark/Hadron Matter\footnote{Invited talk at the second Japanese-Korean
 Joint Workshop on ``Strangeness in Nuclear Matter'', Dec. 6-7, 1993, Seoul,
Korea.}
\normalsize
\vskip 1.0cm
{\bf Tetsuo Hatsuda}\\
\vskip 0.3cm
{\it Institute of Physics, Univ. of Tsukuba, Tsukuba, Ibaraki 305, Japan}\\
\end{center}

\vskip 1cm
\begin{abstract}
A brief review on the recent theoretical progress in hot/dense QCD is given.
 Special emphasis is put on  the non-perturbative aspects of QCD plasma and the
 modification of hadron proporties near the critical temperature of
 chiral transition.
\end{abstract}

\par

 \noindent
\section{Introduction}
\setcounter{equation}{0}
\renewcommand{\theequation}{\arabic{section}.\arabic{equation}}

 Due to the asymptotic freedom, the QCD coupling constant
 $g(\mu)$ decreases logarithmically as one increases the renormalization
 scale $\mu$,
 \beq
{g^2(\mu) \over {4\pi}} \simeq {{12\pi} \over {(33-2N_f)\ln(\mu^2/\Lambda^2)}}
\ \ \ . \eeq
  Now,  $\mu$ should be  chosen to be a typical scale of the system
 to suppress the higher orders in $g$.
 In extremely hot and/or dense QCD medium,
 $\mu$ will thus be proportional to $T$ (temperature) or the
 chemical potential,
 which indicates that the system
 is composed of  weakly interacting  quarks and gluons
  at high $T$ and/or high baryon density $\rho$ (quark-gluon plasma phase).
 On the contrary, quarks and gluons are confined inside mesons and baryons
 at low $T-\rho$ (hadronic phase). Therefore  one expects
 a phase transition between the two phases at certain $T$ and $\rho$.
 This expectation is actually ``proved'' by the numerical simulations
 of QCD formulated on the lattice at finite $T$ \cite{LAT}.
  Experimental efforts to creat and detect the quark-gluon plasma
 in the relativistic heavy-ion collisions have also
 been started at BNL and CERN
 and bigger projects are  planned in these laboratories.
 In this report, I am going to discuss three topics, (i) an intuitive idea
 of the QCD phase transition, (ii) recent theoretical progress in hot/dense
 QCD and (iii) some interesting observables in future experiments.

\newpage
\section{How does the QCD phase transition occur?}
\setcounter{equation}{0}
\renewcommand{\theequation}{\arabic{section}.\arabic{equation}}

Imagine heating  up the  QCD vacuum.  At low $T$, pions (the lightest
 mode in QCD) are thermally excited.  Since the pion has its own
 size (radius $\sim 0.6$fm), the thermal pions start to overlap with each other
 at certain temperature $T_c$ and dissolve
 into a gas of quarks and gluons  above $T_c$.
 One can estimate $T_c$ by identifying
 (pion's volume)$^{-1}$ with the pion number density at $T_c$
  obeying the Bose-Einstein distribution
 $n_{\pi}(T_c)$;
\beq
 n_{\pi}(T_c) = [{4 \over 3} \pi R_{\pi}^3]^{-1},
\eeq
which gives $T_c \sim 200$ MeV.  One can make similar estimate
 for cold but dense matter. The ``critical'' density $\rho_c$
 is evaluated as
\beq
\rho_c =  [{4 \over 3} \pi R_{N}^3]^{-1} \simeq 3 \rho_0,
\eeq
with $\rho_0$ being the normal nuclear matter density (0.17 fm$^{-3}$).

$T_c$ and $\rho_c$ are not too far from the typical hadronic scale, which
 gives us a hope to create quark-gluon plasma
 in the laboratory experiments (such as RHIC at BNL and LHC at CERN).
 Other than the relativistic heavy-ion collisions,
 the phase transtion is believed to take place in the early universe
 (roughly $10^{-5}$ sec after the big bang) and also it
  may exist  in the deep interior of the neutron stars.
  One should note here that we have so far used the word ``phase transtion''
 in a  loose sense.  Whether the transition to the quark-gluon plasma
  is associated with a rapid change
 of a well-defined order parameter or not is still a controversial matter as
 we will see later.

\section{Recent topics in hot/dense QCD}
\setcounter{equation}{0}
\renewcommand{\theequation}{\arabic{section}.\arabic{equation}}

\vspace{0.3cm}
\noindent
{\em 3.1. Multi scale structure at $T>>T_c$}

  It is now widely believed that there exists non-perturbative
physics in the infrared scale even if $T$ is much larger
 than $T_c$ \cite{LI}.  This aspect is summarized as the following three
scale structure:
\beq
 g^2 T \ \ \ \ < \ \ \ \ gT \ \ \ \ < \ \ \ \ T ,
\eeq
with $g$ being the running coupling constant at finite $T$.
 $gT$ is a
 scale of the electric screening mass of gluons and $g^2T$
 is  related to the magnetic
screening mass.  As far as one stays in low orders of $g$, we
 have a consistent perturbation theory for static
and dynamical properties of the system.
   On the other hand, once one tries to go
 to higher orders or to look at
 the system by a probe of energy scale of  $O(g^2T$),
 the perturbation theory starts to break down, e.g.
 at $O(g^6)$ for the free
energy and  at $O(g^4)$ for the magnetic mass \cite{LI}.
 In this case, one has to sum up infinite numbers of diagrams
to get sensible results, which is certainly a non-perturbative task.  $g^2T$
is also related to the coupling constant of 3D QCD as a high $T$
 effective theory of 4D QCD;  the non-perturbative nature of 4D QCD at
  $O(g^2T)$
 has close relation to the confining nature of 3D QCD \cite{IH}.

  As an intuitive tool to understand the
coexistence of non-perturbative ({\bf NP})
and perturbative ({\bf P}) physics at high $T$,
  let us introduce a scale
 $K$ in momentum space which separates the {\bf NP} and {\bf P}. (See  Fig.1.)
  Since the {\bf NP} region is limited in a
finite volume in momentum space,
 it will not affect the bulk properties of the system for high enough
 $T$ where
 the  typical frequency of quarks and gluons is (a few)$\times T \gg K$.
 However, if one probes low frequency region or
 decreases $T$ toward $T_c$, effect of the {\bf NP} region emerges.
 This feature is actually seen in the lattice QCD
 simulations of the energy density ${\cal E}$ and
 pressure ${\cal P}$ \cite{KAS},  and also the hadronic screening
 masses \cite{DK}. Teiji Kunihiro and myself have  predicted
 the non-perturbative phenomena in the hadron spectra above $T_c$
 in the scalar and pseudoscalar channels \cite{HK2}
 before the appearance of the lattice
 data.

\vspace{5cm}

\begin{center}
\footnotesize{Fig. 1: Separation of scales at high $T$.}
\end{center}

\vspace{0.3cm}
\noindent
{\em 3.2. Order of the phase transition}

The determination of the order of the QCD phase transition near $T_c$ is
one of the central issues of the recent lattice QCD study.
  As for the
pure gauge system without dynamical fermion ($m_q=\infty$), the center symmetry
($Z(3)$ in $SU_c(3)$ case) controls the confinement-deconfinement phase
transition.  The effective $Z(3)$ spin model predicts the 1st order transition
and the lattice studies with finite scaling analyses support
this feature \cite{FINITE}.
  Although it is of 1st order, the transition is much weaker than
that seen before on the smaller lattice.
 Once one introduces dynamical
fermions, $Z(3)$ symmetry is explicitly broken. However, as far as $m_q$ is
large enough, one can still study the phase transition based on this
approximate symmetry.
 On the other hand, in the opposite limit where
 $m_q$ is zero,  chiral symmetry instead of $Z(3)$ symmetry takes place to
control the phase transition with
 $\la \bar{q}q \ra_{_T}$ as an order parameter.
 For finite quark masses
 ($m_{u,d} = O(10$MeV) and $m_s=O(200$MeV)),
   chiral symmetry is explicitly broken again,
but one can still study the phase transition based
on the approximate chiral symmetry.
 The recent lattice data \cite{COL} show,
  although the chiral transition near $m_q=0$ seems to
be 1st order, the phase transition is not observed for the realistic
 values of $m_{u,d,s}$. However, this conclusion could be changed
 in the future large scale simulations with finite size scaling analysis.

The order of the chiral transition
 is most relevant to the big-bang nucleosynthesis of
  $^9$Be, $^{10}$B and $^{11}$B \cite{Kajino}.
 The spacial inhomogeneity
 due to the bubble formation during the 1st order chiral transtion
 can creat those relatively heavy elements, while the  standard
 homogeneous model creats
 only 10-100 smaller abandances.

 From the point of view of the relativisitic heavy-ion collisions,
 the precise order of the transion is not much relevant because the
 system size is finite.
 Instead the global behavior of the order parameter and the entropy
 as a function of temperature are rather important for the
 time-evoluion of the system and for related experimental signals
 of the formation of the quark-gluon plasma.

 The energy density and entropy at finite $T$ are known to
  have rapid growth in a  narrow range of temperature
 ($\sim 10$ MeV)  by the numerical simulations on the lattice.
 As for the quark condensates at finite $T$, both lattice QCD and
 QCD effective
 lagrangians  predict (i) a rapid change of  the light quark condensate
  around $T_c$ and (ii) a smooth change of the strange condensate
  across this temperature \cite{HK3}. (See Fig. 2.)
 This might indicate a considerable difference of the behavior of the
 hadronic matter at finite $T$ in the $u,d$ sector and that in the  $s$ sector.

\vspace{6cm}

\begin{center}
\footnotesize{Fig. 2: Light quark condensate (solid line) and the
 strange-quark condensate (dashed line) at finite $T$ in an effective theory
 of QCD \cite{HK3}.}
\end{center}

\newpage
\noindent
{\em 3.3. Dynamical critical phenomena}

Since the quark condensate is a scale dependent quantity and
 is not an observable, one has to look for other physical quantity
   to see the signal of the chiral
 phase transition.  The hadron masses at finite $T$ is one of the
 possible candidates for such quantity.
 In fact, the masses of the light hadrons are essentially determined
 by the quark condensates as QCD sum rules tell us. This suggests
 that the mass shift of hadrons in medium  will be a good measure of the
 partial restoration of chiral symmetry in medium \cite{HK2,BR}.
  There exist similar situations in condensed matter physics: for example,
  the existence of the soft phonon mode
 is an indication that the ground state
  undergoes a structural phase transition.
 In QCD, scalar mesons (fluctuation of the order paramter)
 and the vector mesons (such as $\rho$, $\omega$ and $\phi$)
 are the candidates of the ``soft mode''.
 Su Houng Lee, Yuji Koike and myself have developed a method to calculate
 the mass shift of these mesons (QCD sum rules at finite
 temperature) \cite{KHL1,KHL2}.
 Both in scalar ($\sigma$) and vector channels, one can establish a
 relation between $m(T)$ (meson-mass at finite $T$) and the
 4-quark condensate $\la (\bar{q}q)^2 \ra_{T}$.
\beq
m(T) \leftrightarrow \la (\bar{q}q)^2 \ra_{T}.
\eeq
The r.h.s. of (3.2)
 can be evaluated by the pion gas approximation,
 or hopefully by adopting  the future lattice data.
 We found that, in the scalar and $\rho$ channels, there is a sizable
 softening of the masses near $T_c$. (See, Fig.3.)
 Asakawa and Ko later generalized this approach to the $\phi$ meson
 at finite $T$ by taking into account the thermal strange particles
 to evaluate the r.h.s. of (3.2)  and found a significant softening
 even in this channel \cite{AK}. (See, Fig.4.)

\vspace{7cm}

\begin{center}
\footnotesize{Fig. 3: Light scalar and vector mesons
 at finite $T$ in the QCD sum rules \cite{KHL1,KHL2}.}
\end{center}

\newpage
\ \
\vspace{5cm}

\begin{center}
\footnotesize{Fig. 4: $\phi$ meson
 at finite $T$ in the QCD sum rules \cite{AK}.}
\end{center}

The similar calculation can be also done for the system at
  finite density (but $T=0$) as  has been shown
 by Su Houng Lee and myself \cite{HL}.
  One of the main differences
 of this system from the ($T$$\neq$$0$,$\rho$=0) system is the behavior
 of the quark condensates.   For instance, the light
 quark condensate in the chiral limit
 decreases slowly at low temperature
\beq
\la \bar{q} q \ra_{_T}/ \la \bar{q} q \ra_{_0}
 = 1 - [(N_f^2-1)/N_f]\  T^2/12f_{\pi}^2 + \cdot \cdot \cdot  ,
\eeq
 while it decreases linearly at low density
\beq
\la \bar{q} q \ra_{\rho} / \la \bar{q} q \ra_{_0}
 = 1 - \rho \cdot \Sigma_{\pi N}/f_{\pi}^2m_{\pi}^2
 + \cdot \cdot \cdot  ,
\eeq
 with $\Sigma_{\pi N}=(45 \pm 10)$MeV
 being the $\pi$-N sigma term.
   If one extrapolates the latter formula to normal
nuclear matter density $\rho_0=0.17$fm$^{-3}$,
 the condensate decreases almost 20$-$30\%.
  Accordingly,  rapid
change of hadron
masses at relatively low densities has been predicted \cite{HL}.
 (See, Fig. 5.)  The Walecka model of nuclear matter also predicts
 the similar decrease for the $omega$-meson as was shown recently \cite{WIL}.

\vspace{6cm}

\begin{center}
\footnotesize{Fig. 5: The masses of $\rho$,  $\omega$
 and $\phi$ mesons in nuclear matter \cite{HL}.}
\end{center}

\newpage
Now, what will be the observable consequences of the softening
 phenomenon?   Since the vecotor mesons can decay into lepton pairs,
 $\rho$, $\omega$ and $\phi$ are the best particles to be looked at.
 In the case of $\rho$, the shift of the peak position or the smearing
 the $\rho$-peak could be observed in the future
 relativistic heavy-ion collisions
 (See \cite{KHL2}, \cite{Karsch} and Fig. 6.).
  Also, there is an experimental
 proposal to detect the peak-shift through the lepton pairs \cite{Shimizu}
 where
  $\rho$-mesons are created inside the heavy nuclei using tagged photons
  at INS-ES.
 As for $\phi$, one might be able to see a double peak structure
 in the lepton pair spectrum in the heavy-ion collisions \cite{AK2}.
 This is due to a combined effect of the peak-shift and the
 large entropy jump at the phase transition point.  (See, Fig. 7).
 There is also an experimental
  proposal to detect the modification of the $\phi$-meson through
 lepton and kaon pairs \cite{Enyo}
  where $\phi$-mesons are created in heavy
 nuclei by p-A collisions at KEK-PS.

\vspace{6cm}

\begin{center}
\footnotesize{Fig. 6: The smearing of the $\rho$ peak due to the
 mass shift in the dilepton invariant mass
spectrum in the heavy-ion collisions \cite{Karsch}.  $T_f$ denotes
 the freeze-out temperature and the solid line is a result witout
 the mass shift.}
\end{center}

\vspace{5.5cm}

\begin{center}
\footnotesize{Fig. 7: The dilepton invariant mass spectrum
 at the central rapidity in the heavy-ion collisions \cite{AK2}.
 The solid curve is a result of the initial temperature = 250 MeV
 and the dashed curve is a result without phase transition.
 Right two peaks correspond to $\phi$ (double $\phi$ peak) and the
 peak near 0.8 GeV corresponds to $\omega$.}
\end{center}

Up to this point, we have assumed that the restoration of chiral symmetry
 always takes place in  the central rapidity region of the
 heavy-ion collisions at extreme high energies.
  However, there is another interesting possibility in which a disoriented
 chiral condensate is produced  \cite{BJ}.  This disoriented
 chiral condensate will eventually decays into the vacuum  orientation
 through the emmission of pions. However, because of the event by event
 large fluctuation of the direction of the disorientation, one might
 observe a large isospin fluctuation for the low momentum pions in the
 hadron-hadron and nucleous-nucleous collisions.
 Realistic calculations using the linear sigma-model has been
 started to check whether such phenomenon really occurs or not \cite{other}.
 Also, this might have some  relevance to the cosmic-ray event Centauro.

\section{Summary}

\noindent
(1) {\em Non-perturbative physics above $T_c$}:

 The higher order analyses
 of the perturbation theory have shown that there exists
  non-perturbative
physics even above $T_c$.
 The lattice data
of the free energy  and also the screening mass in the scalar-pseudoscalar
 channel  seem to support the coexistence of perturbative and
 non-perturbative physics above $T_c$.
  There are several ideas about the physics behind them, which include
 the non-perturbative cutoff, the gluon condensate above $T_c$, color singlet
 soft modes above $T_c$ and so on.  Clear physical understanding of
 them are called for.

\vspace{0.3cm}
\noindent
(2) {\em Chiral restoration and its associated phenomena}:

 $\la \bar{q}q \ra $ for light quarks is a good order parameter
 near the chiral limit.
 There is little doubt about the existence of
  chiral transition
 at high $T$ in the chiral limit and probably it is also
 true at high $\rho$.
  Like similar cases in condensed matter and nuclear physics,
 we can expect phenomena associated with this phase
  transition; in particular the
 modification of the elementary modes of excitations (hadrons) in medium.
  QCD sum rules and effective theories predict a sizable softening
 of the scalar and vector mesons in medium.  Furthremore,
 one might be able to see
 such  softning by looking at the dileptons in the future
 laboratory experiments.

\vspace{1cm}

\begin{center}
{\bf Acknowledgements}
\end{center}

I would like to thank Prof. Bhang and the members of the
  organizing committee for the warm hospitality at Seoul National University.

\end{document}